\documentclass[reprint,superscriptaddress,amsmath,amssymb,aps,pra]{revtex4-1}
\usepackage{dsfont}
\usepackage{subcaption}
\usepackage[bottom]{footmisc}
\usepackage{graphicx}
\usepackage{dcolumn}
\usepackage{bm}
\usepackage{xcolor}

\begin{document}
\preprint{APS/123-QED}

\title{Bell-type correlation at quantum phase transitions in spin-1 chain}


\author{Dongkeun Lee}
\affiliation{Department of Physics, Sogang University, 35, Baekbeom-ro, Mapo-gu, Seoul,
04107, Republic of Korea}
\affiliation{Research Institute for Basic Science, Sogang University, 35, Baekbeom-ro,
Mapo-gu, Seoul, 04107, Republic of Korea}

\author{Wonmin Son}
\email{sonwm@physics.org}
\affiliation{Department of Physics, Sogang University, 35, Baekbeom-ro, Mapo-gu, Seoul,
04107, Republic of Korea}

\date{\today}

\begin{abstract}
For the identification of non-trivial quantum phase, we exploit a Bell-type correlation that is applied to the one-dimensional spin-1 XXZ chain. It is found that our generalization of bipartite Bell correlation can take a decomposed form of transverse spin correlation together with high-order terms. The formulation of density-matrix renormalisation group is utilized to obtain the ground state of a given Hamiltonian with non-trivial phase. Subsequently Bell-SLK-type generalized correlation is evaluated through the analysis of the matrix product state. Diverse classes of quantum phase transitions in the spin-1 model are identified precisely through the evaluation of the first and the second moments of the generalized Bell correlations. The role of high-order terms in the criticality has been identified and their physical implications for the quantum phase has been revealed.
\end{abstract}

\maketitle


\section{\label{sec:level1}Introduction}
Quantum correlation, that has no classical counterparts, plays a pivotal role in studying the many-body systems. Investigation of such a complex massive system through the correlation is a major research topic in the statistical and the condensed matter physics as they provide useful view for the quantification of physical properties in a large scaled systems \cite{DeChiara:2018cg}. Especially, the trait of quantum phase and its phase transition driven by the quantum fluctuation in correlation have been attracted much interests throughout the many years as they are closely related to macroscopic quantum phenomena such as superconductivities and topological state of a matter. They are analyzed through the various measures that has been rooted at the quantum information theory.

Quantum entanglement is one of the most crucial concepts that has been extensively studied in the last two decades to comprehend the quantum phase transitions and the exotic quantum properties in condensed matter systems \cite{Amico:2008en}. Concurrence, one of the most used measures for two-qubit entanglement, is exploited to estimate the quantum criticality and, further, its scaling behaviour is also analyzed near the critical point of the one-dimensional XY model \cite{Osterloh:2002gs,Nielsen:2002fs}. Entanglement entropy of a block in various many-body systems also has been discussed during the analysis using the conformal field theory \cite{Eisert:2010hq}. The entanglement entropy demonstrates logarithmic divergence at critical points, whereas it saturates for noncritical systems associated with the area law \cite{Eisert:2010hq,Vidal:2003cn}. Subsequently, entanglement instabilities demonstrates the origin of the quantum criticality precisely and their analytic dynamics as the interaction parameters are modified \cite{Son:2009jc}.

It has been well known that all pure entangled states for two-qubit systems violate CHSH-Bell inequality \cite{Capasso:1973}. On the other hands, the non-local character of mixed entangled states are not easily quantifiable since it does not necessarily violate Bell inequalities \cite{Werner:1989zz}. It means that quantum entanglement is not necessarily equivalent to the concept of Bell non-locality \cite{Werner:1989zz,Augusiak:2015}. Thus, it can be said that the concept of entanglement is closely related to the non-locality while they addresses the different physical quantities in general.

Unlikely to the many-body entanglement, that has been subjected to extensive investigation, the Bell non-locality in many-body systems has been attracted less attention relatively so far. It is because generalization of non-locality is needed to consider all the possible local measurements and locate their non-trivial combinations of the joint probabilities. It is only recent that the non-localities in many-body systems get to make an active progress. At first, Bell correlations for many-body systems are applied to two-site reduced density states which are generated from the ground state in one-dimensional (1D) spin-1/2 XY model \cite{Batle:2010cl} and spin-1/2 XXZ model \cite{Justino:2012iy}. As a result, it has been known that the Bell correlation demonstrates the non-analyticity at the critical point while, due to the monogamy characteristics of the correlation, any bipartite Bell inequality is not violated by the translational invariant many-body systems \cite{deOliveira:2013ga}.

Differently from the bi-partite quantum non-locality, when multipartite correlation is considered, the violation of local realism is possible to be observed, especially in the 1D spin-1/2 model of the nearest-neighbor interactions \cite{Sun:2014cr,Sun:2014gn}. It is also quite recent that non-locality in the Lipkin-Meshkov-Glick (LMG) model can be found through multipartite Mermin inequalities \cite{Bao:2020}. The violation is manifested when the external magnetic field is adjusted. Furthermore, multipartite non-locality, that takes into account one- and two-body correlations imposing on the permutational symmetry, has been found in the Dicke state as well as the ground state of the isotopic LMG model \cite{Tura:2014m,Tura:2015ep}. The violation of the permutational invariant Bell inequality has been experimentally tested by the spin squeezed state that is realized by the large number of atoms in the Bose-Einstein condensate state \cite{Schmied:2016}. It is also found that the maximal violation is observed at the quantum criticality in the Ising model of infinite-range interactions \cite{Piga:2019}.

It is in two-dimensional local Hilbert space that the most Bell correlations are elucidated in aforementioned studies. In other words, Bell correlations in d-dimensional systems of many particles have been rarely investigated, whereas the other measures of quantum information theory, such as fidelity and quantum coherence, are used to characterize the quantum phase transitions (QPTs) many times in spin-1 chains or more \cite{Yang:2008dl,Goli:2013,Malvezzi:2016bq}. In this work, we use a generalized high-order Bell correlation obtained from high-dimensional spin measurements which is proposed by D. Collins {\it et al} \cite{Collins:2002ix} and W. Son {\it et al} \cite{Son:2006,Bae:2018}. Especially, we analyze its Bell correlation in the spin-1 chain to identify its exotic quantum phase near the point of QPTs.

In particular, we choose 1D spin-1 XXZ model with the onsite anisotropy, that is one of the most studied spin-1 chain model. Higher-spin XXZ system have richer quantum phases in various different parameter regimes and certain parts of the system are need to be depicted with more complicated physics like symmetry-protected topological phases \cite{Kjall:2013db}. Since the ground state of spin-1 chains are not possible to be obtained analytically, we utilize numerical techniques, the density-matrix renormalization group (DMRG) method, for the ground state of spin-1 system which are described in matrix product state (MPS) representations. As a result, we found various QPTs that is accurately indicated by analyzing the derivatives of the CGLMP-SLK correlation : FM-to-XY, Haldane-to-AFM, Haldane-to-large-D, and XY1-to-XY2 quantum phase transitions. The first-order QPTs (FM-to-XY transitions) shows discontinuity of the CGLMP-SLK correlation. The Ising-type (Haldane-to-AFM and XY1-to-XY2) QPTs occur where the second derivative of the correlation vanishes. The Haldane-to-large-D transition, also referred to as the Gaussian phase transition, emerges when $\partial\langle\hat{\mathcal{B}}_r\rangle/\partial D=0$. The criticality of the correlation is caused by the fact that both the derivatives of the spin-spin correlations $C^{(1)}_r$ and $C^{(2)}_r$ are simultaneously zero in the criticality. Moreover, we discuss whether violation of the CGLMP-SLK inequality can be detected even in three-dimensional systems.

The outline of this paper is given as following: In Sec.\ref{sec:spin1H}, we starts with the system hamiltonian of the 1D spin-1 XXZ model with the onsite anisotropy. There, we briefly explains its phases at the different parameter regime and their criticalities at the QPTs. In sec.\ref{sec:cglmp}, we presents the Bell correlation in the CGLMP-Bell inequality from the generalized SLK-Mermin-Bell formalism described in an operator form of nonlocal order parameter. In Sec.\ref{sec:mps}, we introduce the matrix product state representations for the ground state of the infinite-size system along with the iDMRG method. Then, using the state obtained, we present our results for the description of Bell correlations in overall parameter space and analysis in the vicinity of the criticality in Sec.\ref{sec:dataanalysis}, and we conclude this paper in Sec.\ref{sec:con}.

\section{The 1D spin-1 XXZ model with the on-site anisotropy}\label{sec:spin1H}
We now start with the description of the system to be analyzed. The Hamiltonian for the
one-dimensional spin-1 XXZ model with the on-site anisotropy is of the form
\begin{align}\label{eq:spin1H}
\hat{H}=\sum_{j} J(\hat{S}^x_j\hat{S}^x_{j+1}+\hat{S}^y_j\hat{S}^y_{j+1}+J_z\hat{S}^z_j
\hat{S}^z_{j+1})+ D \sum_j(\hat{S}^z_j)^2,
\end{align}
where $\hat{S}^{a}_j$ for $a=x,y,z$ denotes the spin-1 operators at site $j$.
In the spin-z basis, the spin-1 operators are given by
\begin{align}
\nonumber
\hat{S}^x=
\frac{1}{\sqrt{2}}&\begin{pmatrix}
0 & 1 & 0\\
1 & 0 & 1\\
0 & 1 & 0
\end{pmatrix},\;\;
\hat{S}^y=
\frac{1}{\sqrt{2}i}\begin{pmatrix}
0 & 1 & 0\\
-1 & 0 & 1\\
0 & -1 & 0
\end{pmatrix},\\
&\;\;\;\;\;\;\;\hat{S}^z=
\begin{pmatrix}
1 & 0 & 0\\
0 & 0 & 0\\
0 & 0 & -1
\end{pmatrix}.
\end{align}
The summation in Eq.\eqref{eq:spin1H} illustrates the one-dimensional spin-1 XXZ chain with
the exchange anisotropy $J_z$ and we set $J=1$ for the unit energy scale.

The second sum in Eq.\eqref{eq:spin1H} describes an on-site anisotropy where the constant D means the uniaxial single-site anisotropy. It has been known that this model can be in the various phases including the antiferromagnetic (AFM) phase, the ferromagnetic (FM) phase, XY phase, the Haldane phase, and the large-D phase. Additionally, the ground state of the system undergoes four types of the QPTs, as varying the parameters $J_z$ and $D$. The ground-state phase diagram of the spin-1 chain Eq.\eqref{eq:spin1H} is investigated in the earlier study \cite{Chen:2003ja}. Here we now briefly demonstrate the characteristics of the different phases and QPTs in the given parameter space.

\paragraph{The AFM and FM phases :}
The antiferromagnetic (AFM) and ferromagnetic (FM) phases describes the system with the magnetic order in it. The fact that all spins in the FM phase are aligned in the same direction gives rise to $\langle\hat{S}^z_j\rangle=\langle\hat{S}^z_{j+1}\rangle$ for all $j$. The spins in FM phase can be quantified by the order parameter $\langle\hat{S}^z_j\rangle$. In the case of AFM phase, spins at the nearest neighbor site $j$ and $j+1$ are aligned in the opposite directions and thereby they satisfies $\langle\hat{S}^z_j\rangle=-\langle\hat{S}^z_{j+1}\rangle$.

\paragraph{The XY phase and BKT-type phase transition :}
The XY phase is that all spins which interact to the nearest neighbors lie in the $xy$ plane and are not allowed to point in the spin-$z$ direction. At a finite temperature $T_c$, Berezinskii, Kosterlitz, and Thouless (BKT) firstly depicted the phase transition without any symmetry breaking in the classical 2D XY model \cite{Berezinskii:1970,Kosterlitz:1973fc,Kosterlitz:1974kk}. While decaying exponentially above the temperature $T_c$, the spin-spin correlation at $T\le T_c$ is expected to decay in a power law, $\langle S^+S^-\rangle \sim r^{-\eta}$ where $0<\eta\le1/4$. These behaviors for the classical XY model still hold for the quantum spin chains \cite{Schulz:1986fd}. The XY phase in the 1D spin-$S$ XXZ model (Eq.\eqref{eq:spin1H} for any spin $S$ and $D=0$) is prevalent for $-1\le J_z \le0$ \cite{Kjall:2013db,Alcaraz:1992dp}: the XY phase of the spin-$1/2$ XXZ chain covers the wider range $-1\le J_z \le1$ while that of the spin-1 system belongs to the range $-1\le J_z\le 0$.  This feature has an influence on the BKT-type quantum phase transition which it takes place at $J_z=0$ for $S=1$ \cite{Kitazawa:1996ea} unlike at $J_z=1$ for $S=1/2$. In Eq.\eqref{eq:spin1H}, two types of the XY phase are manifested \cite{Schulz:1986fd}. In the XY1 phase, the spin-spin correlation $\langle\hat{S}^+\hat{S}^-\rangle$ displays a power-law decay as we mentioned above. For large negative $D$, the XY2 phase appears where $\langle(\hat{S}^+)^2(\hat{S}^-)^2\rangle$ decays following the power law behavior.

\paragraph{The Haldane phase and the large-D phase :}
A variety of phases in many-body systems can be discerned by local parameters that demonstrates the spontaneous symmetry breaking. However, it is the Haldane’s conjecture \cite{Haldane:1983ip,Haldane:1983cl}  that 1D chain of arbitrary integer spin has the gapped phases where any spontaneous symmetry breaking is not manifested. Since then, surveying the characteristics of the Haldane phases becomes significant topic to be investigated.

Later it has been known that the most renown order parameters to capture the Haldane phase is
the string order parameter
\begin{equation}
\mathcal{O}_{str}\equiv \langle \hat{S}^\alpha_i
e^{i\pi\sum_{l=i+1}^{j-1}\hat{S}^\alpha_l}\hat{S}^\alpha_j \rangle,
\end{equation}
for $\alpha\in\{x,y,z\}$ in the presence of the $Z_2\times Z_2$ symmetry \cite{Nijs:1989dp,Kennedy:1992ed}. It is remarkable that this quantity, an expectation value of the nonlocal operator has a nonzero value even in the thermodynamic limit \footnote{Here, a word, nonlocal, is not the same as the Bell nonlocality, a violation of Bell inequality in Sec.\ref{sec:cglmp}. It is also mentioned in a paper, N. Brunner et al., Rev. Mod. Phys. \textbf{86}, 419 (2014)}.

Also, Affleck {\it et al.}  suggested the exactly solvable spin-1 model, called the AKLT state, that exhibits the Haldane phase \cite{Affleck:1987,Affleck:1988hl}. The model is described by the Hamiltonian
\begin{equation}
H=\sum_j \vec{S}_j\cdot\vec{S}_{j+1}+\frac{1}{3}\left(\vec{S}_j\cdot\vec{S}_{j+1}\right)^2
\end{equation}
whose ground state tend to be valanced bonds state with four-fold edge degeneracies. The unexpected properties like edge states and a string order are investigated by using its exact ground state that has non-trivial phases originated from topological order in the system. On the course of the discussion, it is notable that, for the non-trivial phases, bi-quadratic terms are appeared in the interaction Hamiltonian as it is needed to identify the high order moments in the higher spin system.

At the $D>>1$ limit in Eq.\eqref{eq:spin1H}, the ground state becomes large-D phase whose
state takes a rather trivial product form $|00\cdots 0\rangle$. As it is argued in
\cite{Gu:2009ki,Chen:2011iq,Chen:2011}, the state can be identified by the trivial
phase factor from the {\it projective representations of symmetry groups} in the
Schmidt-state basis. Unlikely to the large-D phase, Pollmann et al. demonstrate that the
entanglement spectrum in the Haldane phase of spin-1 chains is doubly degenerated by using
the projective representations in MPS language
\cite{Pollmann:2010ih,Pollmann:2012gb}.

\section{Bell-type correlation from the generalized nonlocality criteria} \label{sec:cglmp}
Bell-type inequalities examine the correlation between two parties of a composite quantum state and determine whether a given quantum state is nonlocal through the violation of local realistic bounds. Clauser-Horne-Shimony-Holt (CHSH) version of Bell inequality is given for the simplest symmetric states where two dichotomic measurements are performed for each of the parties \cite{clauser1969}. If the CHSH correlation admits the local hidden variable model, its absolute value does not
exceed the local realistic bound 2. Taking pure entangled states into account, one can find violation of the CHSH inequality and its maximal violation $2\sqrt{2}$ derived by Tsirelson \cite{Cirel'son:1980}. For the case of two-qubit mixed states, entangled ones do not always violates this inequality \cite{Werner:1989zz}.

When it comes to the systems of arbitrary dimension $d$, the inequality derived by Collins{\it et. al.} provides a generalized version of CHSH inequality \cite{Collins:2002ix}. Further generalization of CHSH for the multiparty $d$-dimesional systems has been made by Son {\it et. al.} which is also consistent to the CGLMP \cite{Son:2006,Bae:2018}. In these scenarios, all the parties are permitted to perform two different measurements at each site with $d$-outcome measurements.

From here, the correlation on the left hand side of the inequality for $d\times d$ system is said to be \textit{CGLMP correlation} as it reduced into the one found by CGMLP \cite{Collins:2002ix}. Our description is made not by set of probabilities but by the form of correlation functions. It can be proved that the representations are equivalent as long as an appropriate collection of the correlations is made through the right choices of weighting coefficients \cite{Bae:2018}.

In the scenario of the CGLMP inequality, there are two observers ($j=1,2$) and each of them has two measurement choices $V_j\in \{A_j, B_j\}$. The measurement operators for the $j$-th party are given as
\begin{equation}\label{eq:observable}
\hat{V}_j\equiv\sum^{d-1}_{\alpha=0}\omega^{\alpha}|\alpha\rangle_{V_j}\langle\alpha|,
\end{equation}
where $d$ is the index of the number of possible outcomes, $\omega^{\alpha}$ is the
eigenvalue, and $|\alpha\rangle_{V_j}$ is the corresponding eigenvector. One observer chooses the Fourier transformed basis and the other observer takes the inverse Fourier basis. The Fourier transformed basis is given as
\begin{align}\label{eq:Fbasis}
|\alpha\rangle_{V_j}=\frac{1}{\sqrt{d}}\sum^{d-1}_{\beta=0}\omega^{-(\alpha+\phi_{V_j})\beta}|\beta\rangle,
\end{align}
where $\omega=\exp(2\pi i/d)$ and $\phi_{V_j}$ denotes a phase shift for a measurement $\hat{V}_j$. Here, the basis $|\beta\rangle$ is nothing but the computational basis like the spin-$z$ basis for $d=2$.

As to be maximal tests, four different measurement choices can be made through four different phase shifts that are $\phi_{A_1}=0$, $\phi_{B_1}=1/2$,$\phi_{A_2}=-1/4$, and $\phi_{B_2}=1/4$. Based upon this expression, CGLMP correlation can be described by the expectation value of the following composite operator,
\begin{align}
\label{eq:cglmp_corr}
\nonumber
\langle\hat{\mathcal{B}}\rangle=\Bigg<
\sum^{d-1}_{n=1}f_n\left(\hat{A}^{n}_{1}+\omega^{n/2}\hat{B}^n_{1}\right)\otimes\left(\hat{A}^{n}_{2}\right.&+\left.\omega^{n/2}\hat{B}^n_{2}\right)^\dagger\Bigg> + c.c.
\end{align}
when the weight function becomes $f_n\equiv\frac{1/2}{d-1}\omega^{\frac{n}{4}}\sec(n\pi/2d)$.
As it is addressed in \cite{Bae:2018}, the weight $f_n$ can be specified for the
convex sum of the probability distributions suggested by Collins {\it et. al.}
\cite{Collins:2002ix}. Bell correlation Eq.\eqref{eq:cglmp_corr} for the case of
$d=2$ is obviously equivalent to one in the CHSH inequality.
Local realistic bounds of the CGLMP correlation for $d=3$ are given by
$-4\le\langle\hat{\mathcal{B}}\rangle\le2$, which can be derived from $f_n$
\cite{Bae:2018}.

It is convenient to investigate the structure of the operator
$\langle\hat{\mathcal{B}}\rangle$ in Eq.\eqref{eq:cglmp_corr} by introducing the $n$-level lowering operator $\hat{J}^n_j\equiv\sum^{d-1}_{\beta=n}|\beta\rangle_j\langle\beta-n|$ for
$j$-th party. From Eq.\eqref{eq:observable} and Eq.\eqref{eq:Fbasis}, one can derive the identities such that
\begin{align}
\frac{1}{2}(\hat{A}^n_1+\omega^{n/2}\hat{B}^n_1)&=\hat{J}^n_1, \\
\frac{1}{2}(\hat{A}^n_2+\omega^{n/2}\hat{B}^n_2)&=\omega^{n/4}\hat{J}^n_.
\end{align}
Thus, the operator $\hat{\mathcal{B}}$ can be simplified as
\begin{align}\label{eq:cglmp_op_J}
	\hat{\mathcal{B}}=\frac{2}{d-1}\sum^{d-1}_{n=1}\sec\left[\frac{n\pi}{2d}\right]
\left(\hat{J}^n_1\otimes\hat{J}^{n\dagger}_2+\hat{J}^{n\dagger}_1\otimes
\hat{J}^{n}_2\right).
\end{align}
This operator holds for the local basis of dimension $d$.
In this work, we analyze the spin-1 chain where the dimension of local basis is set to be three, i.e., $d=3$. When the set of states $\{|\beta\rangle\;|\;\beta=0,1,2\}$ is regarded as the computational basis, the matrix representation of $\hat{J}$ and $\hat{J}^2$ is given by
\begin{align}
	\hat{J}=
	\begin{pmatrix}
		0 & 0 & 0 \\
		1 & 0 & 0 \\
		0 & 1 & 0
	\end{pmatrix}
	\;\;\;\text{and}\;\;\;
	\hat{J}^2=
	\begin{pmatrix}
		0 & 0 & 0 \\
		0 & 0 & 0 \\
		1 & 0 & 0
	\end{pmatrix}
\end{align}
where $\hat{J}\equiv\hat{S}^{-}/\sqrt{2}=(\hat{S}^{x}-i\hat{S}^{y})/\sqrt{2}$ and $\hat{J}^2\equiv\hat{J}\cdot\hat{J}$. The operators shift the higher spin state into the lower spin state in one step and two steps respectively. The choice of measurements $A_i$ and $B_i$ in Eq.(\ref{eq:cglmp_corr}) using the corresponding $\phi_{A_i}$ and $\phi_{B_i}$ lead to the annihilation operators of higher order and they identifies all the matrix elements of the given state. It can be found that the convex sum of the correlation function provides the symmetric selection of relevant probabilities for the correlations in the parties as it is addressed in \cite{Bae:2018}.

\section{MPS Representation and iDMRG method}\label{sec:mps}
In the next step, it is required to investigate the correlation of ground state for the physical properties of many-body systems in Eq. (\ref{eq:spin1H}). We utilize a matrix product state (MPS) representation, which helps to identify the Schmidt
coefficients of the many-body pure state as they are in the product forms of the matrices
\cite{Perez-Garcia:2007:MPS:2011832.2011833,Schollwock:2011gl,Orus:2014ja}.
A quantum state $|\Psi\rangle$ on the 1D systems of length $N$ can be written in canonical
MPS form \cite{Vidal:2003gb}:
\begin{align}\label{eq:canonicalMPS}	
|\Psi\rangle=\sum_{s_1,\cdots,s_N}\Gamma^{[1]}_{s_1}\Lambda^{[1]}\Gamma^{[2]}_{s_2}\cdots\Lambda^{[N-1]}
\Gamma^{[N]}_{s_N}|s_1\cdots s_N\rangle,
\end{align}
where $\Gamma^{[j]}_{s_j}$ is $\chi_{j-1}\times\chi_j$ matrices and $|s_j\rangle$ is the $j$th-site local basis of dimension $d$ with the physical indices $s_j=0, \ldots, d-1$. When the elements of matrices $\Gamma^{[j]}_{s_j}$ for any $s_j$ are expressed as $\Gamma^{[j]}_{s_j;\mu_{j-1}\mu_{j}}$, we call the subscript $\mu$ the bond index and its dimension $\chi$ the bond dimension. The matrices $\Gamma^{[1]}$ and $\Gamma^{[N]}$ at the ends ($j=1, N$) become vectors whose
dimensions are $1\times\chi_1$ and $\chi_{N-1}\times 1$, respectively.
In the case of the periodic boundary conditions, these matrices are of dimensions $\chi_N \times\chi_1$ and $\chi_{N-1} \times \chi_N$
and there is an additional $\Lambda^{[N]}$ connected to them by summing over $\chi_N$ \cite{Verstraete:2004ci}.
The matrices $\Lambda^{[j]}$ are real, diagonal, and of dimension $\chi_{j}\times\chi_{j}$. These matrices should satisfy the relation such that
\begin{align}\label{eq:normalization}
\sum_{s_j}\Gamma^{[j]}_{s_j}\left(\Lambda^{[j]}\right)^{2}\Gamma^{[j]\dagger}_{s_j}=\sum_{s_j}\Gamma^{[j]\dagger}_{s_j}\left(\Lambda^{[j-1]}\right)^2\Gamma^{[j]}_{s_j}=\mathds{1},
\end{align}
which stems from the normalization condition $\langle\psi|\psi\rangle=1$.
In this paper, we are interested in the 1D infinite chains. The infinite matrix product states (iMPS) can be represented by
\begin{align}
	|\Psi\rangle=\sum_{\{\mathbf{s}\}}
\text{Tr}\left[\cdots\Gamma_{s_j}\Lambda\Gamma_{s_{j+1}}\Lambda\cdots\right]|\cdots
s_js_{j+1}\cdots\rangle,
\end{align}
where the superscript $[j]$ on each matrices is omitted since the boundary effect in the thermodynamic limit is insignificant and it is invariant under the translational symmetry.

The diagonal elements $\lambda_{\mu_j}$ in $\Lambda^{[j]}$ are nothing but the Schmidt coefficients and signify the entanglement spectrum of two partitions where the one consist of spins from $1$ to $j$  and the other ranges from $j+1$ to $N$.
The Schmidt state of Eq.\eqref{eq:canonicalMPS} can be written as
\begin{align} |\Psi\rangle=\sum^{\chi_j-1}_{\mu_j=0}\lambda_{\mu_j}|\mu_j\rangle_L|\mu_j\rangle_R,
\end{align}
where the  nonnegative $\lambda_{\mu_j}$ satisfy the property
$\sum_{\mu_j}\lambda^2_{\mu_j}=1$. The Schmidt bases of the left partition and the right partition are given as
\begin{align}\label{eq:LSchmidt}
	|\mu_{j}\rangle_L&=\sum_{s_1,\cdots,s_j}\Gamma^{[1]}_{s_1}\Lambda^{[1]}\cdots
\Gamma^{[j]}_{s_j}|s_1\cdots s_j\rangle,\\ |\mu_{j}\rangle_R&=\sum_{s_{j+1},\cdots,s_N}\Gamma^{[j+1]}_{s_{j+1}}\Lambda^{[j+1]}\cdots
\Gamma^{[N]}_{s_N}|s_{j+1}\cdots s_N\rangle.	
\end{align}
It is notable that the matrix $\Lambda^{[j]}$ signifies the bipartite entanglement of two subsystems $L$ and $R$. This is because the entanglement entropy is expressed as $S(|\Psi\rangle)=-\sum_{\mu_j}\lambda^2_{\mu_j} \log \lambda^2_{\mu_j}$ from its definition.

For the 1D gapped systems, the ground state can be efficiently approximated by this MPS representation \cite{Verstraete:2006ir} and the entanglement entropy of the ground state of a bipartite system $L$ and $R$ is finite according to the area law
\cite{Hastings:2007}. When the system lies in the vicinity of the quantum criticality, the entanglement entropy for the 1D infinite systems diverges logarithmically \cite{Vidal:2003cn}. That is, a lack amount of entanglement that comes from a small bond dimension $\chi$ causes the inefficiency in the MPS representation to describe the ground state at the criticality.
Nevertheless, the scaling properties of finite entanglement at the quantum criticality is suggested based on the conformal field theory \cite{Pollmann:2009ht}.

\begin{figure}
    \centering
    \begin{subfigure}[b]{0.45\textwidth}
       \centering
        \includegraphics[width=\textwidth]{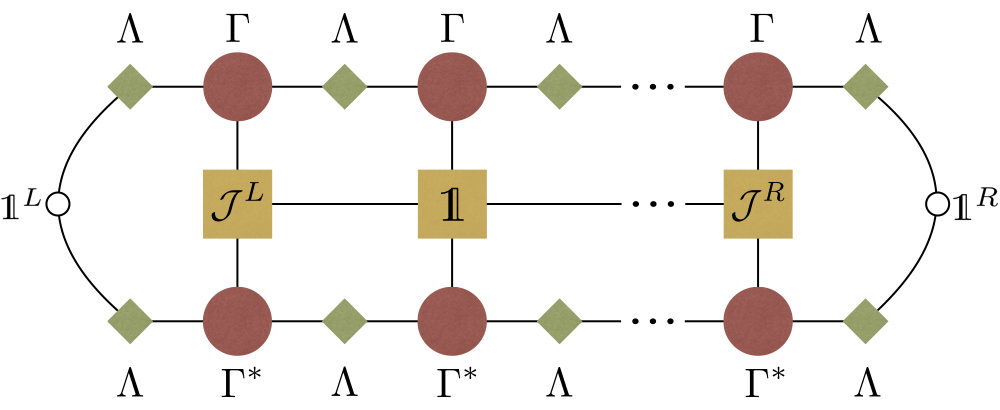}
        {{\small (a) }}
    \end{subfigure}
    \begin{subfigure}[b]{0.45\textwidth}
        \centering
        \includegraphics[width=\textwidth]{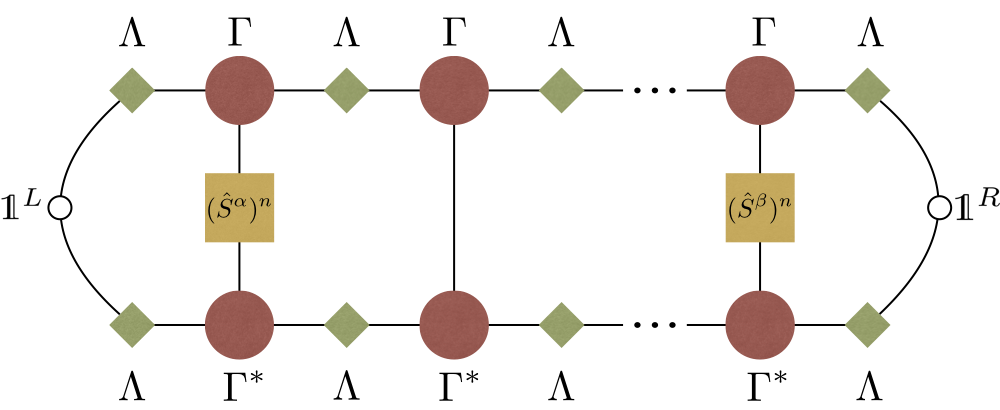}
        {{\small (b) }}
    \end{subfigure}
    \caption[]{\label{fig:expect_cglmp} Graphical descriptions for the expectation value of (a) the operator $\hat{\mathcal{B}}_{i,i+r}$ Eq.\eqref{eq:mpoB} and (b) the spin-spin correlation
    $\langle(\hat{S}^\alpha_i)^n(\hat{S}^\beta_{i+r})^n\rangle$}
\end{figure}

In order to calculate the CGLMP correlation in spin chains which describes the two-body correlation among the many particles, the matrix product states (MPS) representation is applied for the depiction of the ground state and the matrix product operator (MPO) representation for the operator $\hat{\mathcal{B}}$.
Together with the MPS representation for the ground state, the operator $\hat{\mathcal{B}}_{i,i+r}$ between parties $i$ and $i+r$ can be described by the MPO representation that reads
\begin{align}\label{eq:mpoB}	\hat{\mathcal{B}}_{i,i+r}=\hat{\mathcal{J}}^L_i\mathds{1}^{r-1}
\hat{\mathcal{J}}^R_{i+r},
\end{align}
where
\begin{gather*}
	\hat{\mathcal{J}}^L_i =
	\begin{pmatrix}
\sqrt{\frac{2}{\sqrt{3}}}~~\hat{J}_i & \sqrt{\frac{2}{\sqrt{3}}}~~\hat{J}^\dagger_i &
\sqrt{2}~~\hat{J}^2_i &
\sqrt{2}~~\hat{J}^{\dagger 2}_i
	\end{pmatrix}, \\
	\hat{\mathcal{J}}^R_{i+r} =
	\begin{pmatrix}
\sqrt{\frac{2}{\sqrt{3}}}~~\hat{J}^\dagger_{i+r} \\ \sqrt{\frac{2}{\sqrt{3}}}~~\hat{J}_{i+r} \\
\sqrt{2}~~\hat{J}^{\dagger 2}_{i+r} \\
\sqrt{2}~~\hat{J}^2_{i+r}
	\end{pmatrix},
\end{gather*}
and $\mathds{1}$ denotes the identity matrix of dimension $4\times 4$ as shown in Fig.~\ref{fig:expect_cglmp}(a).
Here, $\hat{\mathcal{J}}^L$ and $\hat{\mathcal{J}}^R$ can be obtained by using the method according to Fr\"owis et al. \cite{Frowis:2010ds} and schematic representation of the operator expectation in MPS form is depicted in Fig.\ref{fig:expect_cglmp}.

\paragraph{iDMRG method}
Density-matrix renormalization group (DMRG) method is one of the most widely used numerical methods in one-dimensional systems for infinite size as well as finite size systems\cite{White:1992ie, White:1993fb}. Later, McCulloch showed its infinite version based on the MPO form of the Hamiltonian invariant under the translational symmetry \cite{McCulloch:2008va}. Thus, it is sufficient to just consider an unit cell of two sites in the lattice to find the ground state. The main idea of infinite-size DMRG (iDMRG) method is that (i) the ground state for the
two-site Hamiltonian is obtained, (ii) it is applied to update the environment and
reconstruct the two-site Hamiltonian together with the updated environment, and (iii) this procedure is repeated until the singular-value matrix of the two-site ground state converges. During this process, the Lanczos method is applied to solve the eigenvalue problem for the two-site Hamiltonian and the singular value decomposition is used to obtain the two-site ground state in MPS representation.
The specific procedure of the iDMRG algorithm is well-described in various previous works \cite{Schollwock:2011gl, tenpy}. Here, we simulate the iDMRG method and calculate some expectation values by using TeNPy Library (version 0.4.1) \cite{tenpy}. We choose the bond dimension $\chi=200$ and specially confine a parameter for convergence such that the relative difference of entanglement entropy in each sweep is not over $10^{-7}$.

\section{Quantum criticalities through the correlation in non-locality tests \label{sec:dataanalysis}}
At first, let us present the CGLMP correlation $\langle\hat{\mathcal{B}}_{i,i+r}\rangle$ that is obtained by the expectation value of the operator $\hat{\mathcal{B}}$ for the ground state of Eq.\eqref{eq:spin1H} in the thermodynamic limit. The function is plotted as varying the parameters $J_z$ and $D$ for the spins at different sites distanced $r$.
Here, the class of matrices $\Gamma_{s_j}$ and $\Lambda$ for the ground state can be obtained from the iDMRG method mentioned in Sec.\ref{sec:mps}.
The CGLMP correlation $\langle\hat{\mathcal{B}}_{i,i+r}\rangle$ is calculated through the summation of all tensors $\Gamma$ and $\Lambda$ with $\hat{\mathcal{J}}^L_i$ and $\hat{\mathcal{J}}^R_{i+r}$.
A graphical description for the evaluation of $\langle\hat{\mathcal{B}}_{i,i+r}\rangle$ is given in
Fig.\ref{fig:expect_cglmp}. In the figures, the identity matrix $\mathds{1}^L$ comes from the overlap $\langle\alpha'_{i-1}|\alpha_{i-1}\rangle=\delta^L_{\alpha',\alpha}$ in Eq.\eqref{eq:normalization} and Eq.\eqref{eq:LSchmidt} and the same thing is obtained for $\mathds{1}^R$.

\begin{figure}
    \centering
    \begin{subfigure}[]{0.45\textwidth}
       \centering
        \includegraphics[width=\textwidth]{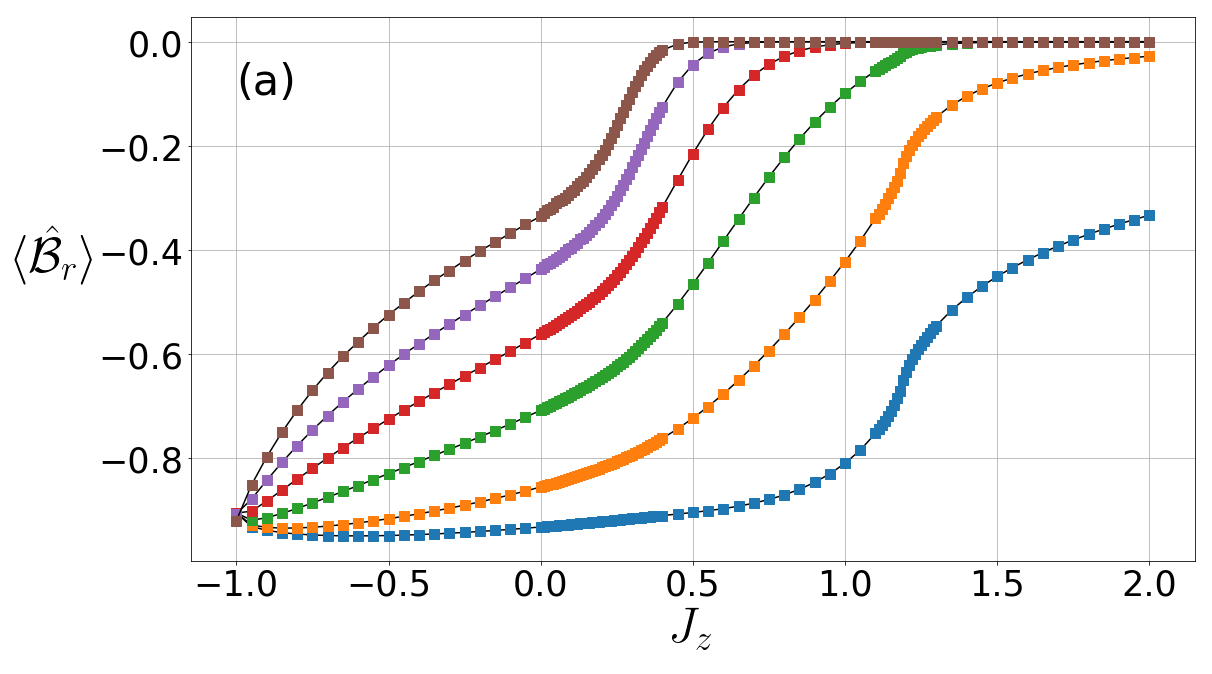}
    \end{subfigure}
    \begin{subfigure}[]{0.45\textwidth}
        \centering
        \includegraphics[width=\textwidth]{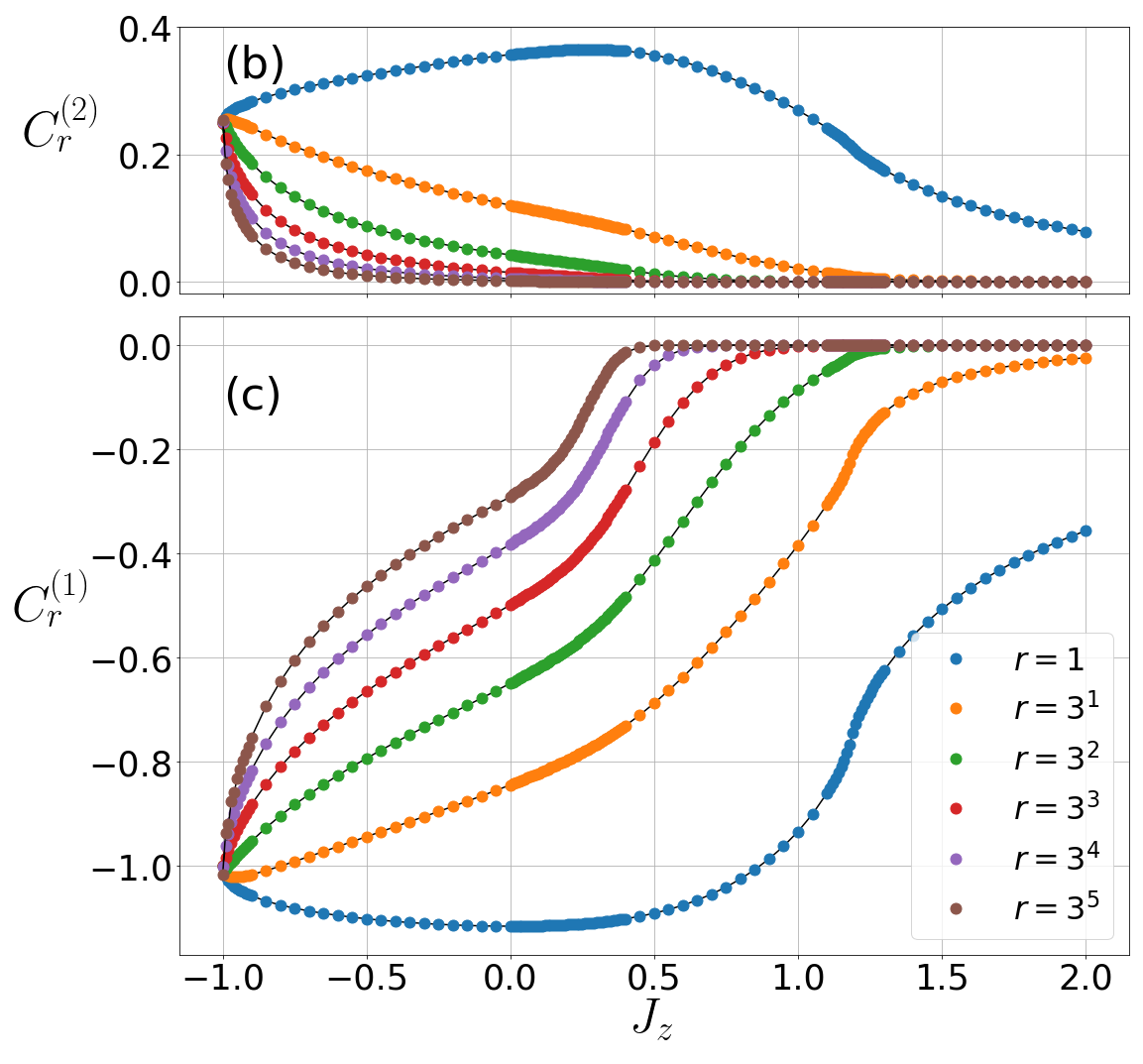}
    \end{subfigure}
    \caption[]{(Color online) (a) the CGLMP correlation $\langle \hat{\mathcal{B}}_{r}
    \rangle$ and (b) the spin-spin correlations $C^{(2)}_{r}$ and (c) $C^{(2)}_{r}$ as
    varying the parameter $J_z$ at $D=0$ for odd distances.}
     \label{fig:Jz}
\end{figure}

\begin{figure*}
    \centering
    \begin{subfigure}[b]{0.48\textwidth}
        \centering
        \includegraphics[width=\textwidth]{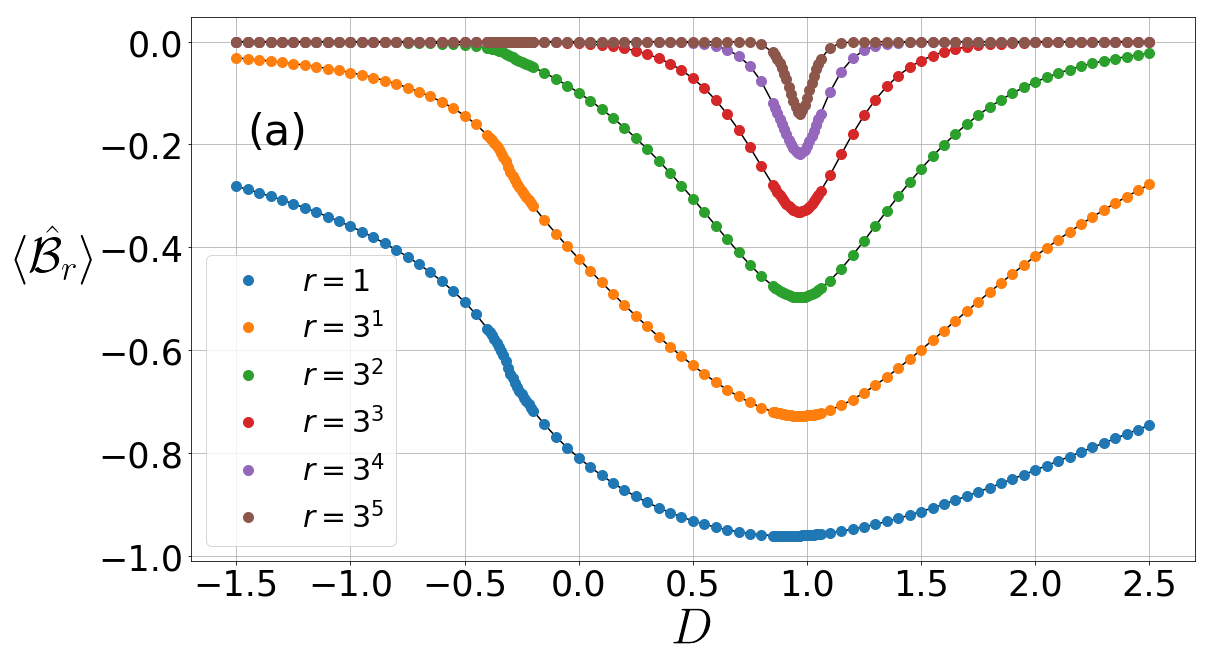}
    \end{subfigure}
    \hfill
    \begin{subfigure}[b]{0.24\textwidth}
        \centering
        \includegraphics[width=\textwidth]{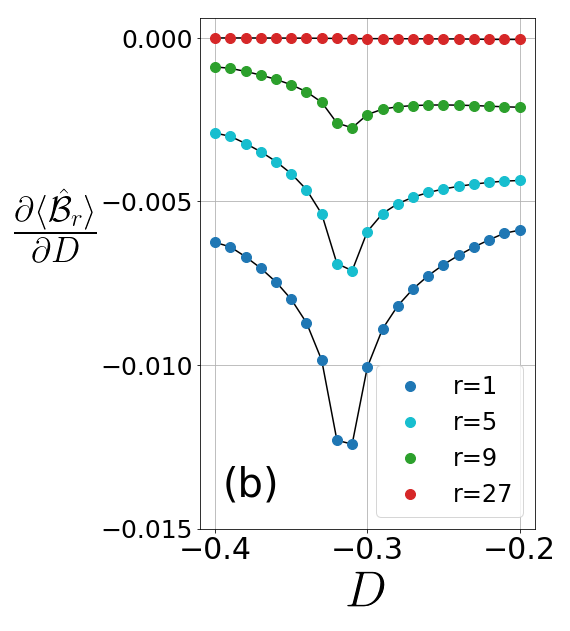}
    \end{subfigure}
    \begin{subfigure}[b]{0.24\textwidth}
        \centering
        \includegraphics[width=\textwidth]{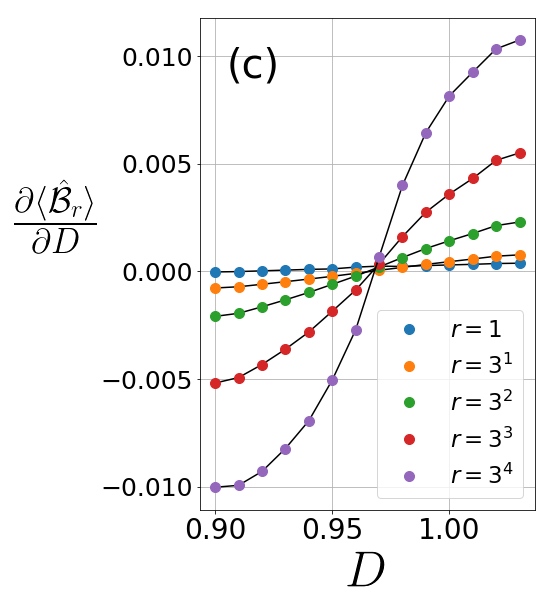}
    \end{subfigure}
    \caption[]{\small (a) the CGLMP correlation $\langle \hat{\mathcal{B}}_{r} \rangle$ and
    the first derivative of $\langle \hat{\mathcal{B}}_{r} \rangle$ with respect to $D$ near
    (b) $D=-0.31$ and (c) $D=0.97 $ as varying the parameter $D$ at $J_z=1$ for odd
    distances.}
    \label{fig:D}
\end{figure*}

\begin{figure*}[t]
    \centering
    \begin{subfigure}[b]{0.36\textwidth}
        \centering
        \includegraphics[width=\textwidth]{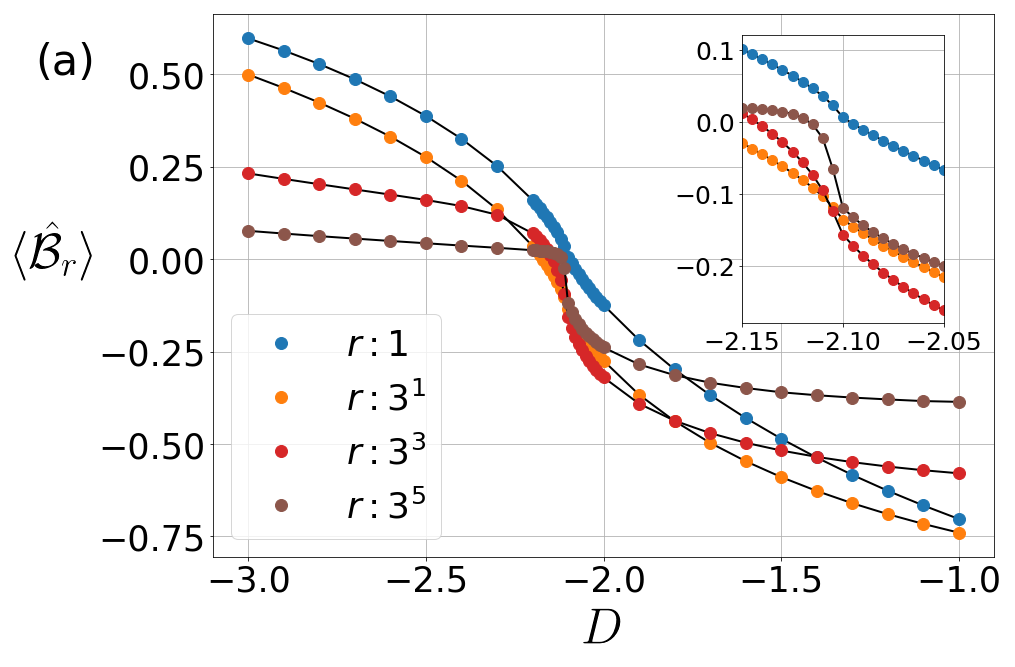}
    \end{subfigure}
    \hskip 2pt plus .01 fill
    \begin{subfigure}[b]{0.37\textwidth}
        \centering
        \includegraphics[width=\textwidth]{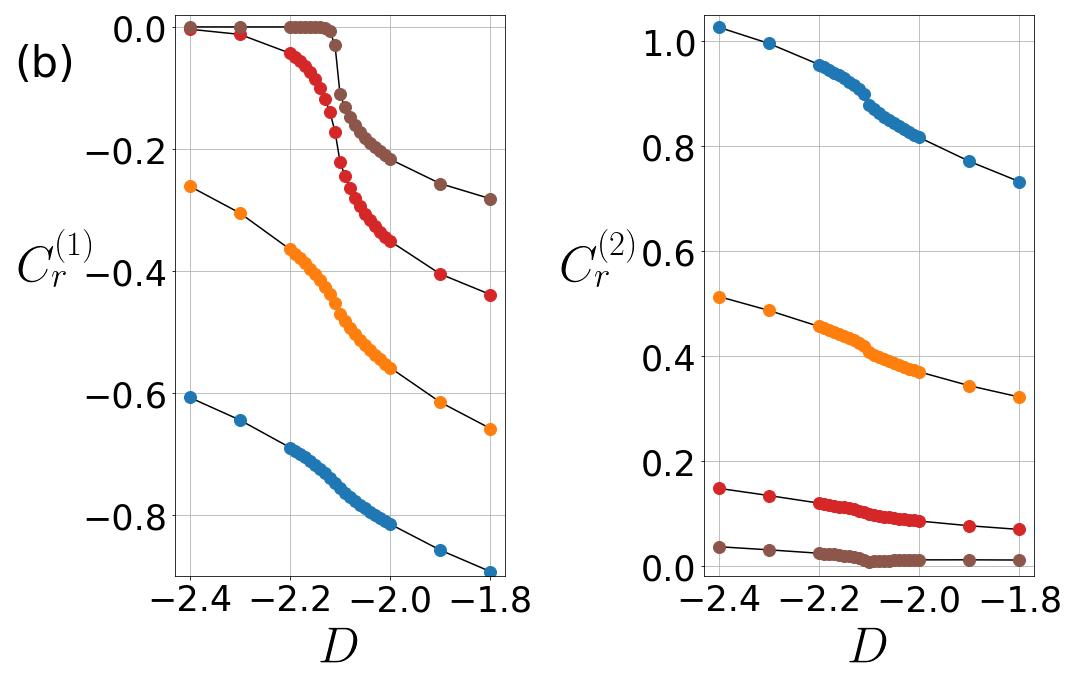}
    \end{subfigure}
    \hskip 2pt plus .01 fill
    \begin{subfigure}[b]{0.21\textwidth}
        \centering
        \includegraphics[width=\textwidth]{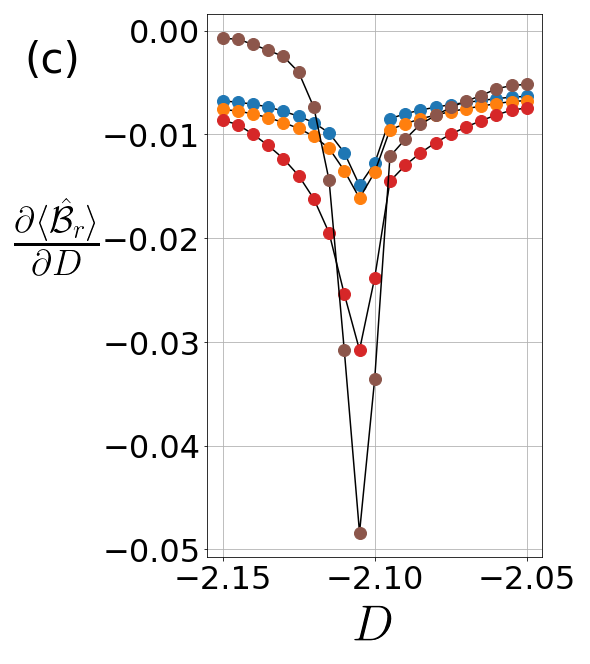}
    \end{subfigure}
    \caption[]{(a) the CGLMP correlation $\langle \hat{\mathcal{B}}_{r} \rangle$, (b) the
    spin-spin correlations $C^{(1)}_{r}$ and $C^{(2)}_{r}$ as varying the parameter $D$, and
    (c) the first derivative of $\langle \hat{\mathcal{B}}_{r} \rangle$ with respect to $D$
    at $J_z=-0.1$ for odd distances $r$.}
    \label{fig:D1}
\end{figure*}

As a result, we here have obtained the characteristics of the CGLMP correlation $\langle\hat{\mathcal{B}}\rangle$ for the particular set of fixed parameters $D=0$, $J_z=1$ and $J_z=-1$. In the parameter region, various phases and QPTs have been identified. The criticalities can be clearly found through the non-analyticity of the correlation function in the first and the second order. Moreover the contribution of higher moment correlation for the criticalities is possible to be discriminated. However, we found that it is not possible to observe the violation of the CGLMP-type inequality in the ground state of Hamiltonian Eq.\eqref{eq:spin1H}. They are presented in Fig.\ref{fig:Jz} (a), Fig.\ref{fig:D} (a) and Fig. \ref{fig:D1} (a). Such non-violation of the CGMLP criteria is consistent to the cases of CHSH in the earlier studies of 1D spin-1/2 XY chain \cite{Batle:2010cl} and XXZ chain \cite{Justino:2012iy}.

The reduced density matrices for two sites in the model of translational symmetry are not possible to violate bipartite Bell inequality. Nevertheless, the CGLMP correlation $\langle \hat{\mathcal{B}}_r \rangle$ can be applied to elucidate the physical properties for the various quantum phases in the system of Eq.\eqref{eq:spin1H}. One of the rationales for the versatile quantification is that Eq.\eqref{eq:cglmp_op_J} for $d=3$ is nothing but the linear combination of spin-spin correlation functions in the higher moments where the $n$-level lowering operator $\hat{J}^n_j$ is regarded as the spin-1 lowering operator $(\hat{S}^-)^n_j$ in this model. It is, thus, given by
\begin{align}\label{eq:Bell_ssc}
	\langle\hat{\mathcal{B}}_{r}\rangle= 2
\left(\frac{1}{\sqrt{3}}\,C^{(1)}_{r}+\frac{1}{2}C^{(2)}_{r}\right),
\end{align}	
where the transverse spin-spin correlations $C^{(1)}_{r}$ and $C^{(2)}_{r}$ are given by
\begin{align}
	C^{(1)}_{r}\equiv \langle \hat{S}^+_i \hat{S}^-_{i+r} \rangle ~~~\mbox{and}~~~
	C^{(2)}_{r}\equiv \langle (\hat{S}^+_i)^2 (\hat{S}^-_{i+r})^2 \rangle.
\end{align}
In Eq.\eqref{eq:Bell_ssc}, we drop the index $i$ because these correlation functions depend not on the site $i$ but on the relative distance $r$ between two spins due to the translational symmetry in the thermodynamic limit.
Based upon the MPS representation, we discuss below how the CGLMP correlation can be used as an indicator of the QPTs by identifying the extremum and inflection points \cite{Wolf:2006cc}. It is also discussed how the Bell-type correlations varies as the distance $r$ changed, especially at the point of QPT, in the following subsections.

\subsection{The CGLMP correlation at $D=0$}
Let us analyze our evaluation of the CGLMP correlation at $D=0$ as varying the anisotropic exchange interaction $J_z$. The CGLMP correlation is zero in the FM phase ($J_z<-1.0$) which is trivial to be illustrated. A discontinuous change of $\langle \hat{\mathcal{B}}_r\rangle$ can be detected at $J_z=-1$
where the first-order QPT occurs at the point. The spins in the AFM phase ($J_z \gtrsim 1.18$) are inclined to locate the spins alternatively in either $|+\rangle$ or $|-\rangle$ site by site. A very large $J_z$ gives rise to nearly vanishing value of the transverse spin-spin correlations and the CGLMP correlation becomes consequently close to zero. The inflection point of the CGLMP correlation can be found at $J_z=1.18$, which turns the AFM phase into different phases. This type of transition is similar to the criticality in the model of two-dimensional Ising interaction. Moreover, the critical point $J_{z}=1.18$ agrees with known results from the previous analysis using the DMRG techniques \cite{Malvezzi:2016bq, HengSu:2012fm, Ueda:2008gn}.

The crucial region for the non-trivial phase is lied at the interval between the FM and the AFM phases. For spin-1/2 system, only XY phase exists in this interval, while there are the XY phase and the Haldane phase for the spin-1 systems. Therefore, more phase transition point exists between the XY phase and the Haldane phase whereas the transition type is going to be the BKT type QPT \cite{Schulz:1986fd}. However, both $C^{(1)}_{r}$ and $C^{(2)}_{r}$ does not show the critical behaviour solely in the region because minimum or the inflection points are not appeared as it can be seen in Fig.\ref{fig:Jz}(b)-(c). Therefore, the CGLMP correlation does not identify the BKT-type QPTs near the point $J_z=0$ at which the QPT in the spin-1 XXZ chain is expected to occur \cite{Chen:2003ja}. Instead of identifying the transition point from the CGLMP correlation directly, the BKT-type QPT can be indicated by an indirect manner such as vanishing string order parameter or a local maximum of the entanglement entropy \cite{HengSu:2012fm,Ueda:2008gn}. In this case, it is demonstrated that the correlation function $C_r^{(1)}$ evaluated from iDMRG method fits a function $ar^{\eta}+b$ near $J_z=0$. Through the extrapolation, the critical exponents $\eta_c=1/4$ of the BKT-type QPT can be found at $J_z=0.02$.

In comparison to the spin-1 case, it is intriguing that BKT-type QPT can be identified by the CHSH correlation in the spin-1/2 XXZ chain \cite{Justino:2012iy}. At $J_z=1$, the spin-1/2 XXZ chain lies at BKT-type QPT and has the SU(2) symmetry. It is possible that the SU(2) symmetry, rather than criticality, is involved in the minimum of the CHSH correlation for $J_z=1$. To be more specific, while $\langle\hat{S}^x_{i}\hat{S}^x_{i+r}\rangle < \langle\hat{S}^z_{i}\hat{S}^z_{i+r}\rangle$ in the AFM phase ($J_z>1$),
$\langle\hat{S}^x_{i}\hat{S}^x_{i+r}\rangle >\langle\hat{S}^z_{i}\hat{S}^z_{i+r}\rangle$ in the XY phase ($J_z<1$).
Due to the SU(2) symmetry, $\langle\hat{S}^x_{i}\hat{S}^x_{i+r}\rangle = \langle\hat{S}^z_{i}\hat{S}^z_{i+r}\rangle$ for $J_z=1$, which affects the maximization of the CHSH correlation with respect to the measurement directions in each phase.

Let us discuss the CGLMP correlation $\langle\hat{\mathcal{B}}_{r}\rangle$ as a function of the distance $r$ between two sites. The decaying behavior of the CGLMP correlations are affected by that of the spin-spin correlations $C^{(1)}_r$ and $C^{(2)}_r$. The spin-spin correlation $C^{(1)}_r$, rather than $C^{(2)}_r$ in Fig.\ref{fig:Jz}(b), dominates $\langle\hat{\mathcal{B}}_{r}\rangle$ without the onsite anisotropy. Then, the CGLMP correlation as a function of $r$ exhibits the even-odd oscillation. Notable feature of the CGLMP correlation is shown in the XY1 phase:
$\langle\hat{\mathcal{B}}_{r}\rangle$ still has a finite value due to a power-law decay of
$C^{(1)}_r$, whereas in other phases it drastically approaches to zero at a large distance $r$ (see Fig.\ref{fig:Jz}).
For the Haldane phase and the AFM phase, the CGLMP correlations $\langle\hat{\mathcal{B}}_{r}\rangle$ are also characterized by an exponential decay, which results from the behavior of the spin-spin correlations $C^{(1)}_r$ and $C^{(2)}_r$.
In Fig.\ref{fig:Jz}(a), CGLMP correlation in the Haldane phase decays less slowly than the AFM phase.
These features corresponds to the fact that the spin-spin correlation function decays exponentially in the AFM phase and one also has an exponential decaying with an additional factor $\sqrt{r}$ in the Haldane phase, i.e. $C^{(1)}_{r}\sim (-1)^rr^{-1/2}\exp(-r/\xi)$ at the Heisenberg point ($J_z=1$ and $D=0$) \cite{Haldane:1983cl,Nomura:1989ka}.
The qualitative forms of the correlation functions  $C^{(1)}_r$ and $C^{(2)}_r$ including various exponents can be expected by using the bosonization technique, an analytical approach based on the effective field theory with defects \cite{Schulz:1986fd} and its exponent $\eta$ and correlation lengths $\xi$ can have precise values by using the DMRG technique \cite{HengSu:2012fm, White:1993ib}.

\subsection{The CGLMP correlation at $J_z=1$}
The Hamiltonian in Eq.\eqref{eq:spin1H} goes through three different phases by tuning the
single-ion anisotropy $D$ at $J_z=1$: the AFM, the Haldane, and the large-D phase.
In addition, there are two types of QPTs that corresponds to an extremum point (either
minimum for an odd $r$ or maximum for an even $r$) and an inflection point of
$\langle\hat{\mathcal{B}}_{r}\rangle$ for each $r$ in Fig.\ref{fig:D}(a).
These critical points are influenced by ones for both correlations $C^{(1)}_r$ and
$C^{(2)}_r$.
In the previous section, the characteristics of the Haldane and the AFM phase have been
discussed and for $J_z=1$ we can also detect the Ising-type QPT at $D_{c_1}\simeq-0.31$ between them,
a peak of $\partial\langle\hat{\mathcal{B}}_{r}\rangle/\partial D$ for small $r$ compared
with the fidelity \cite{Yang:2008dl}.
So, we are going to concentrate on the characteristics of the Haldane phase, and its
relationship with the large-D phase.
It is noteworthy in Fig.\ref{fig:D}(a) that the phase transition between the Haldane phase
and the large-D phase, often called the Gaussian phase transition.

The string order parameter, consisting of spin operators, has a nonzero value only in the
Haldane phase and thereby is widely used to discriminate the Haldane and large-D phase in
this model.
Later, the Haldane phase and the large-D phase are featured by a nontrivial topological phase
and a trivial one under the symmetry protection, respectively
\cite{Gu:2009ki, Pollmann:2012gb}.
It is suggested that the order parameter which reflects the symmetry-protected topological order in the presence of $Z_2\times Z_2$ symmetry becomes -1 for the Haldane and 1 for the large-D phase \cite{Pollmann:2012gb}.
Furthermore, other various studies have revealed the boundary of the Haldane and large-D
phases. The structure for the energy levels has been mainly analyzed to capture the accurate
transition point $D_{c_2}=0.975$ \cite{Ueda:2008gn}.
and also, the behaviors of the ground-energy state have been treated by investigating the
fidelity ($D_{c_2}=0.97$) \cite{Yang:2008dl}, the entanglement entropy
($D_{c_2}=0.96845(8)$) \cite{Hu:2011bo}, which they all use the DMRG method.

In our simulation (see Fig.~\ref{fig:D}(a)), $D_{c_2}$ the minimum points of
$\langle\hat{\mathcal{B}}_{r}\rangle$ which come from those of spin-spin correlations
$C^{(1)}_r$ and $C^{(2)}_r$ are captured around the phase transition between the Haldane and the large-D phase.
For $r=1$, we can find the critical point $D_{c_2}=0.914$ by evaluating the minimum of $\langle\hat{\mathcal{B}}_{r=1}\rangle$.
This critical point moves to $D_{c_2}=0.968$ at $r=81$ by using the linear interpolation near the criticality in Fig.\ref{fig:D}(c).
This critical point converges to $D_{c_2}=0.969$ until the CGLMP correlation decays to zero as the distance $r$ goes to infinity in Fig.\ref{fig:D}(c).
Thus, $\langle\hat{\mathcal{B}}_{r}\rangle$ deserves as a indicator of the QPT for sufficiently distant parties.
It is also remarkable that peak shape in Fig.\ref{fig:D}(a) remains on the criticality even at a large distance $r$.
Specifically, the CGLMP correlation $\langle \hat{\mathcal{B}}_{r} \rangle$ as a function of $r$ decays more slowly in the vicinity of the critical point than in the noncritical region.
It is because the correlation functions $C^{(1)}_r$ and $C^{(2)}_r$ decays as a power law near the criticality, whereas it decays exponentially in the Haldane and the large-D phase \cite{Schulz:1986fd,Nijs:1989dp} and the exponents are well-approximated by using DMRG method \cite{Ueda:2008gn, Hu:2011bo}.

\subsection{The CGLMP correlation at $J_z=-0.1$}

The spin-1 Hamiltonian Eq.\eqref{eq:spin1H} involves two different types of  XY phases, the
XY1 phase and XY2 phases \cite{Schulz:1986fd}. The former which we have been discuss above is located at small $D$ and the latter lies along the large negative $D$.

In order to investigate the CGLMP correlation in two types of XY phases and their transition,
let us take the anisotropy $J_z$ to be $-0.1$ and change the on-site anisotropy $D$.
It is known that for large negative $D$, decay of the correlation function $C^{(2)}_r$ obeys power laws while the correlation $C^{(1)}_r$ decays exponentially, which coincides with the XY2 phase \cite{Schulz:1986fd}.
In Fig.\ref{fig:D1}(c), we can detect the Ising-type criticality between the XY1 and XY2 phases $D_c\simeq -2.10$ by solving $\partial^2 \langle \hat{\mathcal{B}}_{r}
\rangle/\partial D^2=0$.
The farther the distance $r$ is, the deeper the peak of $\partial\langle
\hat{\mathcal{B}}_{r} \rangle/\partial D$ is in the vicinity of this criticality.
This detection quite exactly corresponds to the level crossing point of two energy gaps, one
of which consists of the excitation (or the quantum number) $M^z=\pm 2$ in the XY2 phase and
the other has $M^z=\pm 1$ in the XY1 phase \cite{Chen:2003ja}.

\section{Conclusion\label{sec:con}}
We have investigated the CGLMP correlations in the 1D XXZ model with onsite anisotropy
especially near the quantum phase transitions. In the case of spin-1, the CGLMP correlation obtained from local measurements in the Fourier bases can be interestingly interpreted as a linear combination of the first-order and
second-order transverse spin-spin correlations. By analyzing the derivatives of the CGLMP correlation, one can accurately indicate various QPTs: FM-to-XY, Haldane-to-AFM, Haldane-to-large-D, and XY1-to-XY2 quantum phase transitions.
The first-order QPTs (FM-to-XY transitions) shows discontinuity of the CGLMP correlation. The Ising-type (Haldane-to-AFM and XY1-to-XY2) QPTs occur where the second derivative of the CGLMP correlation vanishes. The Haldane-to-large-D transition, also referred to as the Gaussian phase transition, emerges when $\partial\langle\hat{\mathcal{B}}_r\rangle/\partial D=0$.
The criticality of the CGLMP correlation is caused by the fact that both the derivatives of the spin-spin correlations $C^{(1)}_r$ and $C^{(2)}_r$ are simultaneously zero in the criticality. Moreover, decaying behaviors of the CGLMP correlation with respect to the distance $r$ in each phase can be predicted from the field-theoretical approach \cite{Schulz:1986fd}. However, the BKT-type QPT at $J_z=0$ cannot be indicated by finding critical points of the
derivatives of the CGLMP correlation. This is contrast to the CHSH correlation in spin-1/2 XXZ chain \cite{Justino:2012iy}.

Another crucial result is that there is unfortunately nonviolation of CGLMP inequality in all
phases of spin-1 chain, that is, the CGLMP correlations for the ground states of spin-1 chain
do not surpass the local realistic bounds.
Even though we do not optimize the measurements for the ground states in each phase, this
nonviolation coincides with the one in the 1D spin-1/2 XY chain \cite{Batle:2010cl}
and XXZ chain \cite{Justino:2012iy} invariant under translational symmetry.
The reduced density matrices on two sites which display mixed entangled states are not
sufficient to violate CGLMP inequality in the spin-1 chain, although many-body ground states
are pure and highly entangled in this model.
As the CHSH correlation has monogamy trade-off relation \cite{deOliveira:2013ga},
monogamy of CGLMP correlation can be a candidate to explain the non-violation of CGLMP
inequality in the spin-1 chain.
Therefore, in order to detect nonlocality in this model, one should consider multipartite
Bell correlations for arbitrary dimensions.
This gets a clue from the result that multipartite Bell correlations for spin-1/2 invariant
under the permutational symmetry \cite{Tura:2014m} do not satisfy the local hidden
variable model in the Ising model with long-range interactions
\cite{Schmied:2016,Piga:2019}. We leave that part as a future investigation that is quite consistent
to our current approach demonstrated in this work.

\begin{acknowledgements}
We thank Y. Jo and K. Bae for useful discussions.
This work was supported by the National Research Council of Science and Technology (NST)
(Grant No. CAP-15-08-KRISS), NRF and Samsung.
\end{acknowledgements}

\bibliography{Bell_QPTs_spin1chain}

\end{document}